# THE EXPLANATION OF ENTANGLEMENT IN QUANTUM MECHANICS

H S Perlman

ABSTRACT

It is shown that quantum mechanics is, like thermodynamics, a phenomenological theory i.e., not a causal theory, (not because it is a statistical theory - statistical theories with caused probability distributions can be regarded as causal) but because pure states, i.e. probability distributions of measurement values, cannot inhere in elementary particles and therefore cannot change when their world tubes intersect and hence cannot be regarded as interacting causally. By a causal theory is meant a theory that specifies the changes in time of the states of causally interacting entities in its domain. The areas in quantum mechanics in which causal interactions are relevant include, though not explicitly, measurement and therefore the Born rule, and, explicitly, the unitary Schrödinger time development of states. The Born rule probabilities are shown to refer not to conjoint superpositions of eigenstates but to classical mixtures of mutually exclusive eigenvalues and the Schrödinger time development of states is shown to refer to the time development of the states of non-causally interacting elementary particles and hence cannot be regarded as a causal time development equation, appearances to the contrary. The recognition that quantum mechanics is not a causal theory but a phenomenological theory like thermodynamics does not affect the way it is employed to calculate and predict and hence preserves its empirical success but it does allow a typically simple phenomenological theory explanation of entanglement and other apparently non-local phenomena.

Keywords  Causal interactions . Thermodynamics . Probability distributions . Elementary particles

## 1. TYPES OF PHYSICAL THEORY

All the major physical theories, apart from thermodynamics, including it would seem, quantum mechanics are what intuitively might be called causal theories, that is, theories that include time development equations that specify the effects of causal interactions.

Thermodynamics, the sine qua non of non-causal or phenomenological theories, describes the transitions of the states of the systems it deals with without giving any account of what it is that causes these transitions. It is, for instance, a consequence indirectly of the Zeroth Law of thermodynamics that if two systems at different temperatures inside an isolation enclosure are separated by a diathermal partition, i.e. are in thermal contact, they will ultimately attain the same temperature. This explanation by thermodynamics does not attempt to explain what causes two systems in such contact to behave that way, but contents itself, in conformity with

hperlman@iinet.net.au
School of Physics and Astronomy, Monash University, Clayton, VIC 3800, Australia

the generally accepted notion of an explanation in science, with merely deducing from a very few very widely empirically confirmed basic postulates, or inferences from these, that they do.

What it means for a theory to be causal or to be phenomenological i.e., non-causal, or even what is meant by a causal interaction, is however not precisely clear. The explication of a causal interaction elaborated by Salmon [1] boils down to: a causal interaction occurs when the world lines of two entities intersect and a mark (an alteration which occurs to a characteristic of an entity) is transmitted beyond the spatio-temporal locus of the intersection.

Fair [2] argued that causal interactions between two entities are characterised by the transfer of energy and momentum between the entities.

Dowe [3, 5] proposed, and Salmon [4] subsequently concurred, that a causal interaction occurs when the world-lines of two entities, particles or fields, intersect and there is an exchange of a conserved quantity.

Although the foregoing explications would appear to discern some of the features that characterise causal interactions, it will be necessary to consider more precisely what it is that the theories we regard intuitively as causal in physics have in common.

With respect to statistical theories, a statistical theory is causal if the probability distributions of outcomes predicted by the theory are caused even though individual outcomes can only be predicted with probabilities $< 1$ (Reichenbach [6]). The probability distribution of the outcomes of a toss of a fair die, for instance, is caused even though individual outcomes are uncertain. Whether this is so for quantum mechanics is part of the burden of this paper.

## 2. CAUSAL INTERACTIONS AND CAUSAL THEORIES

Each of the currently generally accepted *intuitively* causal fundamental theories of physics, they include special relativistic classical (i.e. non-quantum) mechanics of particles and continuous matter, solid and fluid, classical electrodynamics (it was always special relativistic), general relativistic gravitation and even quantum mechanics and special relativistic quantum field theory of elementary particles, would appear to be amenable to a Lagrangean formulation (or, derivatively, a Hamitonian canonical formulation) which entails not only the field equations or equations of motion, the differential equations that prescribe the subsequent behaviour of the interacting entities in the domain of the theory but also the changes in measurable constructs alluded to above.

The following closer look at the Lagrangean formulation of physical theories might therefore be of assistance in the explication of causal interactions and causal theories.

It is assumed that there exists a Lagrangean which, in the case of special relativistic classical particles, is a scalar function in Lorentz space-time of the coordinates of the particles and their proper time derivatives and, in the case of fields, is a scalar density function of the field variables and their derivatives in Lorentz space-time or in the case of gravitational fields, in Riemannian space-time.

Hamilton's principle of least action postulates that the way fields and classical particles behave is specified by the requirement that the action, in the case of fields, the scalar space-time volume integral of the Lagrangean density of the fields, is stationary with respect to arbitrary infinitesimal variations in the field variables and in their derivatives that vanish on the boundary hypersurface of a region in space-time, there being zero variations in the coordinates, and by the requirement, that the action, in the case of special relativistic classical particles, the scalar proper time integral of the Lagrangean scalar, is stationary with respect to arbitrary variations in the coordinates of the particles that vanish at the end points.

The said requirement of stationarity entails the Euler-Lagrange equations of motion and field equations for particles and fields. In the case of an isolated system which consists of two what can intuitively be regarded as causally interacting entities, the Lagrangean for the total system is the sum of the Lagrangeans of each of the two otherwise isolated entities and an interaction Lagrangean which is a function of the products of variables pertaining to each of the individual entities.

Since the action functional for any entity or system of entities is a scalar in space-time, its invariance with respect to infinitesimal coordinate transformations, given the Euler-Lagrange conditions for the system, entails that the integral of the divergence of the scalar product of the total energy-momentum tensor of the system and an infinitesimal variation in coordinates over an arbitrary space-time region, vanishes.

This then entails, for a region bounded for convenience by a hypersurface consisting of two space-like planes and a cylindrical hypersurface, parallel to the time axis, out at spatial infinity, that the total energy-momentum vector and the total angular momentum tensor are conserved. Thus, in accord with Noether's theorem, the invariance of the Lagrangean under infinitesimal translations and rotations entails conservation of total energy-momentum and total angular momentum.

And since the interaction term in the total Lagrangean in the Euler-Lagrange equations entails the time development of the system, the energy-momenta and angular momenta of the individual entities will change, that is, there will be a transfer of energy-momentum and angular momentum between the individual entities.

To sum up, it turns out every *intuitively* causal fundamental physical theory does happen to be amenable to a Lagrangean/Hamiltonian formulation, the Euler-Lagrange equations of which, given the appropriate Lagrangean, entail the time development of otherwise isolated interacting systems together with the transfer between them of some of their, conserved in total, energy-momentum and angular momentum. In the case of general relativity, although the total energy-momentum tensor, the total energy-momentum vector and the total angular momentum tensor are actually pseudotensors or pseudovectors, they behave like tensors with respect to linear coordinate transformations and their conservation is independent of reference frame.

It will be apparent therefore that causal interactions between physical entities can indeed be characterised in the sort of way Salmon, Fair, Dowe and others did in the 1980s and 1990s.

To wit, *a causal interaction between say two otherwise isolated physical entities occurs if the world lines or world tubes (to allow for spatial uncertainty and for fields) intersect over a distance and the total energy-momentum, but not the individual energy-momenta, possessed by the two entities is conserved.*

It should perhaps be explained at this point that the ontological status of the entities in physical theories, i.e., particles and fields, whose behaviour a physical theory is concerned with describing in order to explain observations in its domain, are assumed here to be either real or if fictions, fictions which have enough of the attributes of whatever real entities they are surrogates for to explain the empirical success of the theory they inhabit.

Inspection of the total Lagrangean density in the Euler-Lagrange conditions for two otherwise isolated interacting entities, however, makes it clear that while conservation of the total energy momentum but not of the individual energy momenta of the two interacting entities is a sufficient condition for a causal interaction to have occurred, that that does not constitute the "cause" in the causal interaction but is simply a consequence of it. The actual cause, the surrogate for the forces in the collision between two billiard balls, say, is represented in theories amenable to a Lagrangean (or Hamiltonian) formulation by the interaction term in the total Lagrangean (or Hamiltonian). Although force is not an explicit quantity in every physical theory, it persists in its effect in all intuitively causal physical theories in the interaction terms in their total Lagrangeans. If the interaction term in the total Lagrangean of two entities whose world tubes intersect is zero, there is no causal interaction between the entities and the individual energy momentum of each is conserved.

Of course, it is not just the energy-momenta and angular momenta of the individual entities that change in a causal interaction. The Euler-Lagrange equations entail that the variables in the total Lagrangean will also change and since these go to the definition of the states, however these are defined in a particular theory, of the individual entities, they, the states, will necessarily also change. Thus, for instance, if a classical particle undergoes a causal interaction, its state, which is defined by its position and momentum, will change. Likewise, if a quantum particle undergoes a causal interaction, its state, which is defined by the probability distribution of observable values its state vector specifies, will change. In addition to the utilisation of the exchange of energy-momentum as a criterion of a causal interaction, it is therefore possible to employ changes in the states of otherwise isolated entities as a criterion of a causal interaction between them:

*Two otherwise isolated entities interact causally if the states they possess change in the event that their world tubes intersect.*

The explication of a causal interaction in terms of changes of state in individual entities rather than in terms of energy-momentum exchange is entirely consistent with the latter. Not only is this a simpler explication but it also applies outside physics because of the wide applicability of the concept of the state of an entity.

The Euler-Lagrange equations in theories amenable to a Lagrangean (or Hamitonian) formulation specify the time development of the states of entities. If two entities in a Lagrangean/ Hamiltonian formulatable theory interact causally, the interaction Lagrangean/ Hamiltonian term in the Lagrangean/Hamiltonian time development equations will specify the changes in the states of the entities due to the causal interaction of the entities. This leads us to propose here this explication of a causal physical theory:

*A physical theory is a causal theory if it is amenable to a Lagrangean/Hamiltonian formulation in which the Euler Lagrange/canonical equations specify the changes over time in the states possessed by causally interacting entities.*

## 3. CAUSAL INTERACTIONS IN QUANTUM MECHANICS?

For simplicity, we do not here distinguish explicitly distinguish between observables and the self-adjoint operators that represent them and we take the latter to have discrete non-degenerate eigenvalues and normalised eigenvectors.

Pure states in quantum mechanics are understood to refer to individual elementary particles or systems of elementary particles. The pure state of an elementary particle, which is represented in quantum mechanics by a superposition of the eigenvectors of an observable operator in the Hilbert space of the particle, is taken to specify, consistent with the reality of particles and fields posited here, the real physical state of the particle. That is to say, pure states are assumed here to be ontic states (see for instance Harrigan and Spekkens [7].)

The only fundamental assumptions of quantum mechanics concerned with interactions between entities are the Born rule, albeit not explicitly, and, which we will discuss later, the Schrödinger unitary evolution in time of states.

The Born rule, is mostly taken to assert that a measurement of an observable $B$ with eigenvalues $b_n$ on a system, i.e. an elementary particle or a system of elementary particles, in a pure state superposition $|\psi\rangle = \sum_i c_i |b_i\rangle$ of the eigenvectors of $B$ will yield the value $b_i$ with probability, i.e., relative frequency, $P(b_i) = |c_i|^2$ where $c_i = \langle b_i|\Psi\rangle$ and $\sum_i |c_i|^2 = 1$. Born's original formulation [8], and it is often interpreted this way, asserts that $|c_n|^2$ is the probability that a system in the above pure state will be in the state $|b_n\rangle$. Weinberg [9] canvasses both interpretations.

The original interpretation of the Born rule gives rise to the quantum measurement problem, namely, why does a measurement of an observable $B$ on a system in a pure state $|\psi\rangle = \sum_n c_n |b_n\rangle$ yield a single eigenvalue rather than all the eigenvalues to which the eigenvectors in the superposition belong, is generally regarded as one of the fundamental problems in quantum mechanics. That it no longer is will be evident from the following considerations (Perlman [10]) .

The interpretation of the squares $|c_i|^2$ of the moduli of the coefficients $c_i = \langle b_i|\Psi\rangle$ in a pure state superposition of eigenstates $|\Psi\rangle = \sum_i c_i |b_i\rangle$ of the eigenstates $|b_i\rangle$ belonging to an observable $B$ as the probabilities of events (to be identified shortly) together with the closure property $\sum_i |c_i|^2 = \sum_i |\langle b_i|\Psi\rangle|^2 = \langle \psi| \sum_i |b_i\rangle\langle b_i| \psi\rangle = 1$ of complete orthonormal sets of eigenstates belonging to self-adjoint operators in Hilbert space is consistent with the

Kolmogorov additivity axiom $P(\cup_i E_i) = \sum_i P(E_i) = 1$ for *mutually exclusive events*. This entails that the events the Born probabilities $|c_i|^2$ refer to must be mutually exclusive.

The eigenstates $|b_i\rangle$ in a pure state superposition, however, do not constitute a set of mutually exclusive events. To see this, suppose to the contrary that they are mutually exclusive events and that the Born probabilities refer to them in which case the probability that a system in the state $|\Psi\rangle = \sum_i c_i |b_i\rangle$ will be in eigenstate $|b_i\rangle$ is $|c_i|^2$ and the probability that that system, if it is in the eigenstate $|b_i\rangle = \sum_{ij} c_{ij} |a_j\rangle$, is in the eigenstate $|a_j\rangle$ is $|c_{ij}|^2 = \sum_j |\langle a_j|b_i\rangle|^2$. It therefore follows by the said Kolmogorov additivity axiom for mutually exclusive events that the probability $P(|a_j\rangle)$ that the system in the state $|\Psi\rangle$ will be in the eigenstate $|a_j\rangle$ is

$$P(|a_j\rangle) = \sum_i |c_i|^2 |c_{ij}|^2 = \sum_i |c_i c_{ij}|^2. \tag{1}$$

But the straightforward application merely of the closure property of the eigenvectors of self adjoint observables entails $|\Psi\rangle = \sum_i c_i |b_i\rangle = \sum_i c_i \sum_j c_{ij} |a_j\rangle = \sum_j (\sum_i c_i c_{ij}) |a_j\rangle$ and hence

$$P(|a_j\rangle) = \left|\sum_i c_i c_{ij}\right|^2 \tag{2}$$

which contradicts (1). This means that the assumption that the eigenstates in a pure state superposition of eigenstates are mutually exclusive is untenable which, that being an alternative expression of the very measurement problem, is hardly surprising.

If the interpretation of the $|c_i|^2$ as probabilities is to be retained it follows that they must refer to the values $b_i$ yielded by measurements of $B$ and that these values are, there being no impediment to this, mutually exclusive. What this means is that the Born rule needs to be understood to assert that *a measurement of an observable B on a system in the pure state $|\Psi\rangle = \sum_i c_i |b_i\rangle$ will yield the value $b_1$ with relative frequency $|c_1|^2$ or the value $b_2$ with relative frequency $|c_2|^2$ or the value $b_3$ with relative frequency $|c_3|^2$ or...*

This is admittedly not an entirely uncommon interpretation of the Born rule, one which has hitherto been mistakenly regarded as incorrect. But, as just shown, the eigenvalues to which the eigenvectors in a pure state superposition belong, constitute a classical statistical mixture of mutually exclusive values with the Born relative frequencies $|c_i|^2$ in the same way that the numbered faces of a die constitute a classical statistical mixture of mutually exclusive faces with relative frequencies 1/6. And just as the die does not have a particular value, i.e. a particular stationary upward facing numbered face, till it lands with a particular face facing upwards, until it is "measured", so to speak, so an observable pertaining to a system in a pure state does not have a particular value till the observable is measured.

Complementary to the Born rule, the von Neumann projection postulate is usually understood to assert, at least for a certain class of measurements, that if a measurement of an observable $B$ on a system in a state $|\psi\rangle$ yields the value $b_n$ then that state will after the measurement instantly evolve, in a process that is distinct from the Schrödinger unitary evolution of states, into the eigenstate $|b_n\rangle$ belonging to the eigenvalue $b_n$. But given the above revised interpretation of the Born rule and the understanding that the eigenvalues it refers to constitute a classical statistical mixture of mutually exclusive values, it is clear that a measurement will yield a single value in exactly same way that a throw of a die will yield a single face. There is no mysterious non-unitary evolution of state in the quantum case. Pure states, although they are represented mathematically by superpositions of products of probability amplitudes and eigenvectors, function to specify the probability distributions of the eigenvalues to which these eigenvectors belong. A measurement of an observable on a particle in a pure state is a process which selects a value from a classical statistical mixture of mutually exclusive values with a probability distribution specified by the state of the particle. Once a measurement yields a value, the mathematical representation of the state of the particle changes from a superposition of eigenstates to the eigenstate belonging to that value. No Schrödinger or mysterious non- Schrödinger evolution of state is involved. Nor does it have to be postulated. The von Neumann projection postulate is entailed by the above revised version of the Born rule.

Both the revised version of the Born rule and the recognition that pure states, notwithstanding that they are represented mathematically in quantum mechanics by superpositions of eigenvectors, specify probability distributions not of the eigenvectors but of the eigenvalues to which the eigenvectors belong, allow the immediate resolution of the Schrödinger cat problem (Schrödinger [11]), the purpose of which was to ridicule the interpretation of the of the the state of an entity in quantum mechanics as a superposition of states given the uniqueness of its state macroscopically. Schrödinger considers a cat which is enclosed for an hour in a chamber in which there is a device which has a probability ½ of killing the cat. According to the Born rule as usually understood, the cat is during that hour, assuming cats are within the province of quantum mechanics, in the *conjoint* superposition of dead and live states $|\Psi\rangle = \sqrt{½}|dead\ cat\rangle + \sqrt{½}|live\ cat\rangle$ , an embarrassing state for every expert in quantum mechanics let alone every self-respecting cat. But given the revised version of the Born rule, the cat should be understood to be in something like the state $|\Psi\rangle = \sqrt{½}|zero\ pulse\rangle + \sqrt{½}|normal\ pulse\rangle$, pulse being co-opted as an observable, which specifies that a measurement of the cat's pulse during that hour will yield either zero with probability ½ *or* normal with probability ½ . It is still a problem, but only for that cat.

The fundamental assumption in quantum mechanics explicitly concerned with changes in the states of entities in the event of their interaction is the Schrödinger unitary development of state equation. It will be instructive to compare the way causal theories, quantum mechanics and phenomenological theories explain how the states of the entities in their domain change.

As an example of a phenomenological theory explanation we consider how thermodynamics explains the decrease in temperature produced by the adiabatic demagnetization of a paramagnetic solid. The thermodynamic variables that specify the state of a paramagnetic solid are the magnetic field intensity $H$ , the magnetization $M$ , and the absolute temperature $T$ and the equation of state of a paramagnetic solid above temperatures of 1 K, Curie's law, is $M = C\ \frac{H}{T}$ where $C$ is a constant.

It is shown in thermodynamics (it applies to any simple system given the appropriate state variables) that
$$TdS = C_H dT + T \left(\frac{\partial M}{\partial T}\right)_H dH \qquad (3)$$

where $dS = \frac{dQ}{T}$ denotes entropy change. Given the paramagnetic system under consideration here, $C_H = \left(\frac{dQ}{dT}\right)_H$ is the specific heat of the paramagnet at constant $H$.

If the paramagnet, after undergoing isothermal magnetisation, suffers an adiabatic, i.e. $dQ = 0$ and therefore $dS = 0$, decrease $dH$ in the magnetic field it is being subjected to, then, from (3)
$$dT = -\frac{T}{C_H}\left(\frac{\partial M}{\partial T}\right)_H dH \quad . \qquad (4)$$
Since $T$ and $C_{\mathcal{H}}$ can be regarded as approximately constant, the change $dT$, due to $dH$, being small compared to $T$, and since $\left(\frac{\partial M}{\partial T}\right)_H$ is negative (increases in temperature decrease the magnetization of a paramagnet), it follows from (4) that the temperature of a paramagnet in a magnetic field decreases if the magnetic field is decreased adiabatically.

As an example of an explanation by what is a causal theory according to the explications argued earlier, we consider how special relativistic classical mechanics explains the change in state that a charged particle undergoes in the presence of an electromagnetic field.
The Lagrangean for a charged particle being acted on by an external electromagnetic field is
$$\mathcal{L}_{Tot} = \mathcal{L}_{Part} + \mathcal{L}_{Int} \quad , \qquad (5)$$
where $A^\mu$ are the four-potentials specifying the electromagnetic field and $A^{\mu,\nu} = g^{\nu\alpha}\frac{\partial A^\mu}{\partial x^\alpha}$, where
$$\mathcal{L}_{Part} = \tfrac{1}{2}\, m_0\, c\, g_{\mu\nu}\, \dot{x}^\mu(s)\, \dot{x}^\nu(s) \qquad (6)$$
and $m_0$ is the proper mass of the particle, $e$ the charge, $c$ the velocity of light, $s = ct$, and $t$ the proper time, $x^\mu(s)$ and $\dot{x}^\mu(s) = \frac{dx^\mu}{ds}$ the position and the velocity of the particle respectively, and where $\mathcal{L}_{Int} = {}^e/_c\, g_{\mu\nu} A^\nu(x)\, \dot{x}^\mu(s).$ (7)
With this Lagrangean, the Euler-Lagrange equations of motion yield the Lorentz force on a charged particle in an electromagnetic field,
$$m_0\, c\, \ddot{x}^\mu = {}^e/_c\, \dot{x}_\nu\, (A^{\nu,\mu} - A^{\mu,\nu}) \qquad (8)$$
which, given the initial position and energy-momentum i.e., initial state, entails its state at any subsequent time.

As an example of the way quantum mechanics explains how states change with time we consider the interaction, between times $t_0$ and $t$, of two otherwise isolated particles 1 and 2, initially at time $t_0$, in the states $|\theta^1\rangle = \sum_i c_i^1 |a_i\rangle$ and $|\chi^2\rangle = \sum_j c_j^2 |b_j\rangle$ where $|a_i\rangle$ and $|b_j\rangle$ are eigenvectors belonging to the eigenvalues $a_i$ and $b_j$ of the observables $A$ and $B$ pertaining to particles 1 and 2 respectively. If $H^{Tot} = H^1 + H^2 + H^{Int}$ where $H^{Tot}$, $H^1$, $H^2$ and $H^{Int}$ are, respectively, the Hamiltonians of the composite system, particle 1, particle 2, and the interaction Hamiltonian and if effectively $H^{Tot} \approx H^{Int}$ over the interval $t - t_0$ where $H^{Int} = H^{Int}(ZB)$, $Z$ being an observable pertaining to particle 1, the initial state of the total system, at $t_0$,
$$|\psi^{12}(t_0)\rangle = \sum_{ij} c_i^1\, c_j^2 |a_i\rangle|b_j\rangle \qquad (9)$$
develops due to the interaction between the particles into the no longer separable state
$$|\psi^{12}(t)\rangle = \sum_{ijk} c_i^1\, c_j^2\, c_{ik}\, U^{Int}(z_k b_j, t, t_0)|z_k\rangle|b_j\rangle \qquad (10)$$

at $t$, where $c_i^1 = \langle a_i | \theta^1 \rangle$, $c_j^2 = \langle b_j | \chi^2 \rangle$, $c_{ik} = \langle z_k | a_i \rangle$ and $U^{Int} = exp \frac{-i}{\hbar} H^{Int} (t - t_0)$ is the Schrödinger unitary time development operator.

This entails that the probability $\acute{P}(a_w)$ at $t$ that a measurement of the observable $A$ on particle 1 will yield the value $a_w$ at $t$, $\acute{P}(a_w) = \langle a_w | \rho^1 | a_w \rangle$, where $\rho^1 = Tr_2 \rho^{12}$ and $\rho^{12} = |\psi^{12}(t)\rangle\langle\psi^{12}(t)|$, is

$$\acute{P}(a_w) = \sum_{skn} d_n^{1*} \, d_k^1 |c_s^2|^2 \, c_{wk}^* \, c_{wn} \, U^{Int\dagger}(z_n b_s, t, t_0) \, U^{Int}(z_k b_s, t, t_0), \quad (11)$$

where $d_k^1 = \langle z_k | \theta^1 \rangle$.

The interaction between particles 1 and 2 causes, according to (10), or at least appears to cause, a change in the state of the system from $|\psi^{12}(t_0)\rangle$ to $|\psi^{12}(t)\rangle$, which is to say, a change in the probability distribution of the values yielded by measurements of $A$ on particle 1 from $P(a_w) = |c_w^1|^2$ at $t_0$ to $\acute{P}(a_w)$, given by (11), at $t$.

*We recall that the explanation (i.e. retrodiction) or prediction of a particular phenomenon in the domain of a theory is generally understood to consist (in what is called the D-N or deductive-nomological conception of scientific explanation) in the precise entailment of that phenomenon by the theory given particular conditions such as boundary or initial conditions.*

Special relativistic classical mechanics explains that a given initial state, i.e., a given position and momentum, possessed by a charged particle will change to a subsequent state specified by the Euler-Lagrange time development equation in the event that it interacts causally with an electromagnetic field.

Phenomenological thermodynamics explains that the temperature of a paramagnet subjected to a magnetic field decreases by a specified amount if the magnetic field is decreased adiabatically by a specified amount. Thermodynamics is silent about why the demagnetization *causes* the change in temperature.

Quantum mechanics explains that two isolated elementary particles in a given initial composite state $|\psi^{12}(t_0)\rangle$ will change into a subsequent state $|\psi^{12}(t)\rangle$ that is entailed by the apparently causal unitary time development of states equation in the event that the two particles undergo an apparently causal interaction and hence that the probability distribution of the values that measurements of observable $A$ on particle 1 will yield will change from $P(a_w)$ at $t_0$ to $\acute{P}(a_w)$ at $t$.

There is however a problem. When the state of a classical particle changes in a classical mechanics interaction, the momentum the particle possesses changes from $p_i$ at an initial $t$ to $\acute{p}_i$ at a subsequent $\acute{t}$ and we know how a classical particle possesses, physically possesses, pre-measurement, momentum, i.e., how momentum physically inheres in the particle.

Particles in pure states in quantum mechanics do not possess pre-measurement values of observables. An ever increasing number of no-go theorems, all amply confirmed experimentally, attest to that. We mention only a few: the corollary to Gleason's theorem [12] for which there is a stand-alone proof by Bell [13]; Clauser and Horne [14], following Bell [15, 16]; Greenberger, Horne and Zeilinger [17] and Greenberger, Horne, Shimony and Zeilinger [18]: Hardy [19] and Kochen and Specker [20].)

But although elementary particles in pure states do not possess pre-measurement values of observables, they do possess, in the same way that classical particles possess individual pre-measurement values, the pre-measurement probability distributions of values specified by their states. The pure state of an elementary particle, which is represented in quantum mechanics by a superposition of eigenvectors of an observable, specifies according to the Born rule the relative frequency distribution of eigenvalues that measurements of the observable will yield. But question is this. How does a relative frequency distribution of values physically inhere in a structureless particle? How does a structureless elementary particle physically possess a probability distribution of values ?

We know how a probability distribution of throw outcomes equal to 1/6 inheres in a fair die. It, the probability distribution, inheres in the cubic shape and uniform density of the die, i.e., in its structure. A die will have an entirely different probability distribution of outcomes if the uniformity of its density or of its shape is altered. But what actually changes physically in a structureless elementary particle when its state, i.e., the probability distribution of measurement values specified by that state, changes?

It is difficult to avoid the conclusion that the concept of an elementary particle as a structureless and point-like physical particle is inconsistent with its physical possession of a probability distribution of measurement outcomes.

Nor do elementary particles as conceptualised in quantum field theory seem capable of possessing probability distributions of measurement outcomes. The Dirac field, for instance, is specified by non-self-adjoint operators in Hilbert space which obey the Dirac field equation and particular anti-commutation relations. The products of the operator coefficients and their adjoints (the destruction and creation operators) in the Fourier transforms of the Dirac field operators are commuting self-adjoint operators (the occupation number operators) with integral eigenvalues. These eigenvalues are taken to designate electron and positron numbers with particular momenta and spins, and the simultaneous eigenvectors of the occupation number operators belonging to these eigenvalues constitute basis vectors in terms of which vectors specifying the states of electrons and positrons can be expressed. The effect of a particular creation or destruction operator on any one of these eigenvectors, and ultimately on the vacuum state, is to increase or decrease by one the particular occupation  number eigenvalue to which the eigenvector belongs and thereby increase or decrease by one the number of those particular electrons and positrons. Just as in the case of elementary particles in non-field quantum mechanics, there seems to be no capacity in these particles, that is. in the excitations in the Dirac field from which they emerge, to possess probability distributions.

Despite this, quantum mechanics attributes states, which is to say, pre-measurement probability distributions of values, to elementary particles.

But if the state an elementary particle is in is unable to physically inhere in the particle, that is, if there is no physical connection between the relative frequency distribution of values specified by the state and the particle, then the relative frequency distribution of values yielded by measurements cannot be regarded as having actually been *caused* by the particle being in that state. Nor, if relative frequencies distributions of values specified by states cannot physically inhere in elementary particles, can changes in those relative frequencies in the event of the intersection of particle world tubes be said to constitute *causal* interactions.

And yet the probability distribution of the results of measurements of an observable on an elementary particle in a pure state is entirely in accord empirically with the probability distribution specified by the state notwithstanding that probability distributions cannot inhere in elementary particles and therefore cannot be said to cause the observed distribution of results. So how to explain the total empirical success of quantum mechanics?

Is quantum mechanics, despite its ostensibly causal time development of states equation, actually a phenomenological theory, that is, a non-causal theory like thermodynamics? To see that it is indeed a phenomenological theory it might help to recall what we mean by a causal interaction and by a causal physical theory.
Two otherwise isolated entities *interact causally* if the states they possess change in the event that their world tubes intersect.
A physical theory is *a causal theory* if it is amenable to a Lagrangean/Hamiltonian formulation in which the Euler Lagrange/ canonical equations specify the changes in the states possessed by *causally interacting* entities over time.
(And by "possess" we mean here possess physically in the same way that momentum is possessed by or physically inheres in a classical particle.)

The problem in the case of quantum mechanics is this: for particles to interact causally, their pure states, that is, the probability distributions of measurement values they physically possess, have to change when their world tubes intersect. But since elementary particles cannot physically *possess* probability distributions of values, they do not possess what needs to change and hence cannot be regarded as interacting causally any more than could the interactions in classical mechanics between particles be regarded as causal were it, suppose, impossible because of their constitution, for them to physically possess momenta.

For a physical theory to be a causal theory it has to be amenable to a Lagrangean/ Hamiltonian formulation in which the Euler Lagrange/canonical equations specify the changes in time in the states of causally interacting entities. The problem in the case of quantum mechanics is that it is amenable to a Hamiltonian formulation in which the Schrödinger canonical equation analogue specifies the changes in time not of states, i.e. probability distributions, which elementary particles are unable to physically possess but of states, i.e., probability distributions, which quantum mechanics imputes to them on the basis of the outcomes of measurements subsequent to their preparation.

The essential elements of the thesis being argued here are these; for there to be a causal interaction between two entities, it is necessary for the entities to possess the measurable constructs that define their states (in quantum mechanics, probability distributions of measurement values) and for these states to change in the event that their world tubes intersect. And since the entities in quantum mechanics are elementary particles which, being structureless, cannot physically possess probability distributions, that is, cannot physically possess states that need to change in the event that their world tubes intersect for their intersection to constitute a causal interaction, it follows that the Schrödinger time development equation does not specify the changes in the states of *causally* interacting particles and so quantum mechanics is not a causal theory according to what we mean by a causal theory.

To deny that elementary particles *possess* states is not to deny that they are *in* states imputed to them by the theory on the basis of the way these are prepared, that is, that they behave *as if* they possess states, and that being in those states entails according to the theory (but does not cause) particular frequency distributions of measurement outcomes. Nor is denying that elementary particles possess states to deny that the Schrödinger time development equation specifies the changes in the states the theory imputes to world tube intersecting particles just as, for instance, equation (4) specifies the cooling of a paramagnetic solid due to adiabatic demagnetization in thermodynamics, which is to say, descriptively, albeit quantitatively and precisely. Both theories are phenomenological and therefore do not invoke causal interactions. The difference is that the quantum mechanics looks deceptively like it does.

None of the foregoing deprives quantum mechanics of its empirical success, but it does deprive it of its status as a causal theory. That quantum mechanics must be considered a phenomenological theory, it should be understood, is not because, as a phenomenological theory, it avoids some of the problems that encumber it as a causal theory. Quantum mechanics is of necessity a phenomenological theory like thermodynamics because it is not a causal theory.

It will be observed that quantum mechanics and thermodynamics are phenomenological theories for different reasons. Although the former invokes a time development equation, the entities in its domain do not physically possess the states they are in. Thermodynamics, on the other hand, does not invoke a time development equation but the entities it refers to do physically possess the states *they* are in.

The upshot of the recognition that quantum mechanics is a phenomenological theory and not a causal theory, is threefold:
(i) all the postulates and inferences from them, the formalism of quantum mechanics, and the way it is employed to calculate and predict how atomic and subatomic systems behave remain completely unaffected. This includes the employment of the time development of states equation to describe changes in imputed states in the event of, albeit non-causal, interactions.
(ii) as a consequence of (i), all the empirically confirmed explanations and predictions of quantum mechanics remain completely unaffected,
(iii) but some of the foundational problems that have plagued quantum mechanics for most of its history are resolved.

Take the phenomenon of entanglement. How do you explain that space-like separated measurements of the components of the spin observables in the z direction of two distant non-world tubes intersecting spin ½ particles 1 and 2 in the entangled state

$$|\psi\rangle_{EPR} = \sqrt{½} \, ( \, | + 1\rangle_z^1 \; | - 1\rangle_z^2 \; - \; | - 1\rangle_z^1 \; | + 1\rangle_z^2 \, ) \tag{12}$$

always yield the values +1 and −1, respectively, or −1 and +1, respectively, given that a plethora of no-go theorems rule out the assumption of pre-measurement existing values and special relativity rules out superluminal propagation of cause as possible explanations?

The quantum mechanical explanation of the above correlations, that is, their entailment by the theory given particular conditions, once it is recognised that quantum mechanics is a phenomenological and not a causal theory, is simply this: *The Born rule in quantum mechanics entails that space-like separated measurements of spin in the z direction on the two particles in the entangled state* (12) *will always yield one or other of the pairs of values +1, −1 or −1, +1.*

But is this not merely what quantum mechanics predicts? How can it be an explanation? Well, yes, it is what quantum mechanics predicts. But it is also what constitutes the explanation of a phenomenon in a phenomenological theory. True, it is not a causal explanation in a causal theory. A causal explanation in a causal theory would invoke a causal interaction and the operation of a time development of states equation. A causal explanation would tell you what causes the correlations. But, if as has been argued, quantum mechanics is a phenomenological theory, not a causal theory, its predictions are, just as are those of thermodynamics, also explanations, albeit not causal explanations.

That the explanations not just of entanglement but of all phenomena that quantum mechanics offers are non causal need not of course preclude their possible causal explanation by some future causal theory that incorporates the empirical success of quantum mechanics in an entirely different conceptual framework.

## 4. TO SUM UP

It is shown that quantum mechanics is not, despite appearances to the contrary, a causal theory but a phenomenological theory like thermodynamics.

In respect of the concepts of a causal interaction and of a causal physical theory, two otherwise isolated entities, particles or fields are considered here to *interact causally* if the states, which are assumed to refer to their real physical states, however these are defined in a particular theory, which they possess (i.e., which inhere in them in the same way that momentum, say, is possessed by or physically inheres in a classical particle or a probability distribution of throw outcomes is possessed by or inheres in a die) change in the event that their world-tubes intersect. And a physical theory is considered to be a *causal theory* if it is amenable to a Lagrangean/Hamiltonian formulation in which the Euler-Lagrange/canonical equations specify the changes in time in the states possessed by causally-interacting entities. According to this explication, all the fundamental theories of physics, with the exception of thermodynamics, would appear to be causal theories. (Statistical theories with caused probability distributions can be regarded as causal.) A non-causal theory like thermodynamics is called a phenomenological theory.

Tangentially but relevantly, it is shown that due to the closure property of complete orthonormal sets of eigenvectors of self-adjoint operators and the Kolmogorov additivity axiom for mutually exclusive events, the probabilities in the Born rule in quantum mechanics refer to mutually exclusive events. Further, it is shown that the eigenvectors in a pure state superposition $|\Psi\rangle = \sum_i c_i |b_i\rangle$ of the eigenvectors of an observable $B$ (with non-degenerate discrete eigenvalues for simplicity) *do not* constitute a set of mutually exclusive events. It follows, firstly, that if the interpretation of the $|c_i|^2$ as probabilities is to be retained, they must refer not to the eigenvectors $|b_i\rangle$ but to the eigenvalues $b_i$ to which the eigenvectors belong, and secondly, that the eigenvalues constitute a classical mutually exclusive set of events.

The Born rule therefore must necessarily be taken to assert that $|c_i|^2$ is the probability tha a measurement of $B$ on a system in the state $|\Psi\rangle = \sum_i c_i |b_i\rangle$ will yield the value $b_i$ with probability $|c_i|^2$ *or* the value $b_k$ with probability $|c_k|^2$, an interpretation hitherto regarded as incorrect.

This revised version of the Born rule provides the solution to the measurement problem in quantum mechanics. A measurement of an observable on an entity in a pure state superposition of the eigenvectors of an observable will yield a single eigenvalue rather than all the eigenvalues to which the eigenvectors in the state belong simply because the eigenvalues constitute a set of mutually exclusive values and a measurement will yield just one of them.

That quantum mechanics is not a causal theory is entailed by essentially two considerations. The first is that pure states of elementary particles specify relative frequency distributions of values yielded by measurements. The second is that elementary particles are generally conceived as structureless and point-like. It follows from these that relative frquency distributions of values cannot physically inhere in elementary particles. Elementary particles can physically possess mass and charge and the values of other defining constructs in the same way that classical particles can, but they cannot physically possess probability distributions of measurement values because unlike, say, dice, they are structureless. No physical changes occur *in* an elementary particle when its state changes from $|\Psi\rangle = \sum_i c_i |b_i\rangle$ to say $|\acute{\Psi}\rangle = \sum_i \acute{c}_i |b_i\rangle$.

Quantum mechanics *imputes* states to particles based on the outcomes of measurements subsequent to their preparation or on their unitary time development in the event of their interaction with another particle. But since elementary particles cannot physically possess states (i.e. probability distributions cannot inhere in them) the states they do not possess cannot change in the event that their world-tubes intersect and the particles therefore cannot be said to actually interact causally in quantum mechanics. And if that is so then quantum mechanics is a not a causal theory but a phenomenological theory like thermodynamics.

That quantum mechanics is a phenomenological theory does not affect the way the theory is employed to calculate and predict how atomic and subatomic systems behave, it being understood that the Schrödinger time development equation specifies changes in imputed states and therefore changes in the probability distributions of measurement values that are consequent on non-causal interactions But what recognition that it is phenomenological does do is to allow the explanation of some otherwise unexplainable phenomena like entanglement. Entanglement is explained within, i.e. by, quantum mechanics in the following way: *Given particles in an entangled state, the Born rule asserts that measurements on them will yield correlated outcomes.* Quantum mechanics will not explain what *causes* this to happen anymore than would that other phenomenological theory, thermodynamics, were entanglement in its domain.

References

1. Salmon, W. C.: Scientific Explanation and the Causal Structure of the World. Princeton University Press. Princeton N.J. (1984).
2. Fair, D.: Causation and the flow of energy. Erkenntnis. **14**, 219-250 (1994).
3. Dowe.: Wesley Salmon's process theory of causality and the conserved quantity theory. Philosophy of Science **59**, 195-216 (1992).
4. Salmon, W. C.: Causality without counterfactuals. Philosophy of Science **61**, 297-312 (1994).
5. Dowe, P.: Causality and conserved quantities: a reply to Salmon. Philosophy of Science **62**, 321-333 (1995)                                           .
6. Reichenbach, H.: The Direction of Time. Berkley-Los Angeles: University of California Press. (1956).
7. Harrigan, N., Spekkens, R. W.: Einstein incompleteness and the epistemic view of quantum states. Found. Phys. **40,** 125-157, (2010).
8. Born, M.: Zur quantenmechanik der stossvorgange. Z. Phys. **38**, 803-807 (1926).
9. Weinberg, S.: Lectures on Quantum Mechanics. 2$^{nd}$ edn. Cambridge University Press. Cambridge (2015).
10. Perlman, H. S.: Why quantum measurements yield single values. Found. Phys. **51**, 26 (2021). https://doi.org/10.1007 /s10701-021-00440-1
11. Schrödinger, E.: Die gegenwertige situation in der quantenmechanik. Naturewissenschaften. **23**, 807-812, 823-828, 844-849 (1935) Reprinted in Wheeler, J. A., Jurek, W. H.: Quantum Theory and Measurement. Princeton University Press. Princeton N.J. (1983).
12. Gleason, A. M.: Measures on the closed sub-spaces of Hilbert spaces. J. Math. and Mech. **6**, 885-893 (1957).
13. Bell, J. S.: On the problem of hidden variables in quantum mechanics. Rev. Mod. Phys. **38**, 447-452 (1966) .
14. Clauser, J. F., Horne, M. A.: Experimental consequences of objective local theories. Phys. Rev. D. **10**, 526-535 (1974).
15. Bell, J. S.: On the Einstein-Podolsky-Rosen paradox. Physics **1**, 195-200 (1964).
16. Bell, J. S., Introduction to the hidden variable question. In d'Espagnet, B. (ed.)

Foundations of Quantum Mechanics. Academic Press. New York. (1971).

17.  Greenberger, D.M., Horne, M. A., Zeilnger, A.: Going beyond Bell's theorem. In Kafatos, M. (ed.) Bell's Theorem, Quantum Theory, and Conceptions of the Universe. Kluwer. Dordrecht. (1989).

18.  Greenberger, D.M., Horne, M. A., Shimony, A., Zeilnger, A.: Bell's theorem without inequalities. Am. J. Phys. **58**, 1131-1143 (1990).

19.  Hardy, L.: Nonlocality for two particles without inequalities for almost all entangled states. Phys. Rev. Lett. **71**, 1665-1668 (1993).

20.  Kochen, K., Specker, E.: The problem of hidden variables in quantum mechanics. J. Math. and Mech. 17, 59-87 (1967).